\documentclass[aps,prl,nopacs,twocolumn]{revtex4}
\usepackage{amsmath,graphicx,epsfig,amssymb,subfigure,times,dsfont}
\usepackage[usenames]{color}
\usepackage{bbold}

\newcommand {\Fig}[1] {Fig.~\ref{#1}}
\newcommand {\Eqn}[1] {Eq.~(\ref{#1})}

\begin{document}

\title{Unbiased Monte Carlo for the age of tensor networks}

\author{Andrew~J.~Ferris}
\affiliation{ICFO---Institut de Ciencies Fotoniques, Parc Mediterrani de la Tecnologia, 08860 Barcelona, Spain}

\date{\today}

\begin{abstract}
A new unbiased Monte Carlo technique called Tensor Network Monte Carlo (TNMC) is introduced based on sampling all possible renormalizations (or course-grainings) of tensor networks, in this case matrix-product states. Tensor networks are a natural language for expressing a wide range of discrete physical and statistical problems, such as classical and quantum systems on a lattice at thermal equilibrium. By simultaneously sampling multiple degrees of freedom associated with each bond of the tensor network (and its renormalized form), we can achieve unprecedented low levels of statistical fluctuations which simultaneously parallel the impressive accuracy scaling of tensor networks while avoiding completely the variational bias inherent to those techniques, even with small bond dimensions. The resulting technique is essentially an aggressive multi-sampling technique that can account for the great majority of the partition function \emph{in a single sample}. The method is quite general and can be combined with a variety of tensor renormalization techniques appropriate to different geometries and dimensionalities.
\end{abstract}

\pacs{???}

\maketitle

\section{Introduction}

From the mid-1900's onwards, we have seen numerical methods grow from a useful calculational tool into an entirely new way of performing science, somewhere between the traditional theoretical and experimental disciplines. Particularly within physics, chemistry, mathematics and related fields, highly accurate numerical results derived from first principles can now often be compared favourably to experiment where (often perturbative) theoretical approaches fail --- or even used \emph{in lieu of} experiment to simulate theories under difficult to achieve conditions, such as modelling the internals of subatomic particles.

Arguably one of the, if not \emph{the}, most successful and pervasive numerical techniques has been the Monte Carlo method~\cite{Metropolis1949}, which uses statistical sampling to approximate complicated calculations. In their seminal 1953 paper~\cite{Metropolis1953}, Metropolis \emph{et al} boldly claim to introduce ``a general method... of calculating the properties of any substance considered to be composed of individual interacting'' classical particles. Since then, the Metropolis algorithm in particular and Monte Carlo techniques in general have been applied with acclaimed success to fields as far flung as finance, biology and engineering. 

In the intervening years, many new Monte Carlo algorithms have been introduced to extend the method and improve the accuracy obtained for a given amount of computational effort. Improvements in accuracy can roughly be separated into two categories: improving the generation of independent statistical samples, and increasing the accuracy of a single sample. In the former, Markov chain algorithms such as the Metropolis algorithm generate new samples by incremental changes to the earlier samples, introducing a degree of autocorrelation between the samples. This is detrimental because the statistical error $\Delta E$ of a measured quantity $E$ generically decreases with the number of \emph{completely independent} samples $N$ as
\begin{equation}
    \Delta E = \sqrt{\frac{\mathrm{Var}\bigl[E\bigr]}{N}} \label{error}
\end{equation}
Autocorrelation decreases the number of effectively independent samples, reducing the denominator. Important techniques to reduce autocorrelation include cluster updates to avoid the well-known slowing down in critical problems~\cite{Swendsen1987,Wolff1989}, loop updates for quantum or topological problems~\cite{Evertz1993,Evertz2003}, and worm algorithms to sample across off-diagonal quantum observables~\cite{Prokofev1998}.

The second class of improvements act to decrease the numerator in \Eqn{error} by decreasing the sample-to-sample variance of the observable $E$. Monte Carlo methods work to approximate a very large sum over many variables according to their probabilities. Importance sampling, which modifies the probability distribution, is an effective way to ``flatten'' the distribution of the summands so that each contribution has similar values. In practice, importance sampling can only reduce $\mathrm{Var}[E]$ only up until a point, for instance to the variance observed in the ensemble of physical realizations of the system. 

A second path to reducing the variance is by, within each sample, explicitly summing a tractable fraction of the large sum that Monte Carlo is attempting to estimate. Sometimes this is achieved relatively trivially, by exploiting symmetries such as translational invariance, or it may involve moderately challenging calculations at each step to perform the partial summation. In any case, a single sample may be said to be \emph{effectively} performing multiple samples of the sum --- a process (when implemented explicitly) referred to as multi-sampling~\cite{Sankowski2003}. In this work, a new, generic multi-sampling technique is introduced that can lead to massive reductions in $\mathrm{Var}[E]$ for a wide range of classical and quantum statistical problems on a lattice. \emph{In the best-case scenario, the error may decrease exponentially with increasing computational effort}, as opposed to the slower $N^{-1/2}$ scaling of traditional Monte Carlo.

This new method is based on modern techniques for renormalizing (by which I mean `approximately summing') tensor networks. Tensor networks are an extremely powerful language for expressing problems (such as the partition function of classical or quantum systems) and methods for their contraction (which is, for all intents and purposes, a sum over many terms). Tensor network calculations came to prominence with the advent of White's Density Matrix Renormalization Group (DMRG) in 1992~\cite{White1992}, famous for its extraordinary accuracy in solving one-dimensional quantum systems, and which is intimately connected with a tensor decomposition known as matrix product states (MPS)~\cite{Ostlund1995} or sometimes tensor trains. MPS and DMRG techniques can similarly be applied to the contraction of 2D lattices of tensors, by utilizing an MPS to represent a boundary state~\cite{Orus2008} or modifying the original DMRG algorithm to simultaneously approximate the four quadrants of the lattice~\cite{Nishino1997}. In all these cases, the techniques brought an unprecedented level of accuracy for simulating strongly-correlated systems.

From here, tensor network research has been focused on two complimentary efforts. The first, which I call the `variational' approach, attempts to capture a representation of a (typically, quantum) state using a tensor network decomposition that is efficient to store and manipulate on a computer and which can be optimized variationally to obtain the state with lowest energy --- the ground state. Examples of such states include MPS, tree tensor network states~\cite{Shi2006,Tagliacozzo2009}, projected entangled-pair states (PEPS, for 2D quantum systems)~\cite{Verstraete2004,Jordan2008,Gu2008}, the multi-scale entanglement renormalization ansatz (MERA, which can represent scale-invariant states)~\cite{Vidal2007b,Vidal2008} and branching unitary structures for metallic phases~\cite{Evenbly2014b,Ferris2014}. In each of these, the `bond-dimension' or dimensionality of the indices in the tensor network state $D$ is free to be determined by the programmer, resulting in a control which can increase accuracy at the expense of increased computational cost. At the low-end of $D=1$ we recover the set of product states, and thus mean-field theory, but as $D$ increases the accuracy may increase rapidly, even exponentially. As a result, these variational approaches have proven very competitive, even for the study of challenging strongly-correlated (including fermionic) 2D quantum systems (e.g. \cite{Yan2010,Corboz2014,Corboz2014b}).

The second approach, which I call the `renormalization' approach and was mentioned above, seeks not to approximate the state or wavefunction of the system but rather directly approximates the calculation in question, for example, the calculation of a partition function (and thus free energy) or another observable of a statistical problem which is naturally expressed as the contraction of a tensor network. Examples include boundary-MPS, corner transfer matrix, the tensor renormalization group (TRG) of Levin \& Nave~\cite{Levin2006} and later `higher-order' tensor renormalization group (HOTRG)~\cite{Xie2012}, as well as the recent (and broadly named) tensor network renormalization (TNR) of Evenbly \& Vidal~\cite{Evenbly2014}. In each method, one attempts to solve the system by successively approximating larger and larger blocks. In each method, one has control of the bond-dimension $D$ which represents number of degrees of freedom retained in each course-grained (or `renormalized') block, while the remaining degrees of freedom are discarded, introducing a degree of error. And like the variational approaches, in each method, the accuracy has been observed to improve extremely rapidly with $D$, by being able to account for the vast majority of the partition function.

The marriage of tensor networks with Monte Carlo is an idea that has been attempted before. The appeal is obvious --- the combination of the rapidly-improving accuracy of tensor networks with the speed, parallelism and unbiased nature of Monte Carlo would breed a best-of-both-worlds calculation technique. Apart from~\cite{Wouters2014} (which introduces a variational fixed-node or constrained-path approximation), the attempts to date have focussed on variational Monte Carlo, where Monte Carlo is used to sample and update variational tensor network states~\cite{Sandvik2007,Wang2011,Ferris2012,Ferris2012b,Iblisdir2014,Sikora2015}. Unfortunately, optimizing a tensor network state with Monte Carlo estimates has proven challenging due to the combination of statistical errors with extremely stiff optimization problems. The resulting algorithm rather combined the \emph{worst} aspects of variational states and large statistical errors!

The silver-lining of these experiments has been the ability to extract observables from tensor network states which would otherwise be too computationally demanding to calculate --- for example the energy of PEPS with large bond dimension~\cite{Wang2011}. % or Renyi entropies of quasi-2D systems. 
In fact, it partly is this limited success which has motivated this work. While optimization problems involving many variables (thousands to millions) proved unstable to statistical errors, the calculation of a single number such as the norm of a wavefunction is a well-controlled problem.

Thus we now turn to applying Monte Carlo to tensor renormalization techniques where the goal is the extraction of a single (or just a few) numbers from the contraction of a large tensor network. In fact, standard Monte Carlo techniques are applied exactly in this way --- by choosing a single configuration for each bond of the tensor network representation of the partition function, the result decomposes into a simple product of numbers. Further, we will see standard Monte Carlo is equivalent to the $D=1$ version of the technique introduced here!

Where I extend this technique is by multi-sampling: keeping $D > 1$ of the degrees of freedom associated with each bond. Normally, this would result in an intractable tensor network contraction, but the ideas of tensor renormalization are used to calculate each sample efficiently.  At each iteration, course graining is performed, and a subspace of dimension $D$ of the bonds associated with the enlarged block are selected, before repeating. Importance sampling is performed in an optimized basis, and degrees of freedom are selected according to their individual weights which are correlated with the magnitude of their contribution to the total partition function. What is observed is that each sample can account for the great majority of the partition function, and the sample-to-sample variance decreased extremely rapidly with increased $D$, resulting in errors bars of extremely small magnitude which are rare even in Monte Carlo studies of very simple systems. In the systems studied here, it was even possible to forgo Markov chain sampling and generate the samples directly (via `perfect' sampling), eliminating all autocorrelation effects. The Monte Carlo averages are unbiased and do not display the variational error typical to tensor network techniques. The method is seen to combine the best aspects of Monte Carlo and tensor network calculations, and the improvement over standard Monte Carlo has a certain analogy with the improvement of variational tensor network states over mean-field theory.

The purpose of this paper is to outline this new Monte Carlo algorithm based on sampling over the space of tensor renormalizations (or truncations, or projections) called Tensor Network Monte Carlo (TNMC), and to demonstrate numerically that (a) the TNMC algorithm inherits extremely small sample-to-sample variance from the legendary precision of tensor network methods and (b) the variational bias of the tensor network technique is removed by the unbiased sampling. The algorithms and example systems are chosen to be among the simplest possible for clarity: matrix-product states applied to integrable 2D classical systems at thermal equilibrium. However, it should be kept in mind that the core of the Monte Carlo sampling algorithm can easily be applied to more advanced tensor network renormalization schemes, and that the techniques can be applied to both out-of-equilibrium and/or quantum problems in $d$-dimensions by contraction of a $(d+1)$-dimensional tensor network, both of which are examined in the Discussion.

\section{Monte Carlo Algorithm}

\subsection{Boundary MPS calculations}

I begin by explaining a simple tensor network algorithm for contracting (approximately) a two-dimensional tensor network on a square lattice. This type of contraction problem occurs in many scenarios, including solving 2D classical systems at thermal equilibrium and 1D quantum lattice systems via the path-integral formulation, as well as certain statistical/combinatorical problems. The core ideas can be applied to different lattice geometries and higher-dimensional problems.

\begin{figure*}[t]
\centering
\includegraphics[width=2.05\columnwidth]{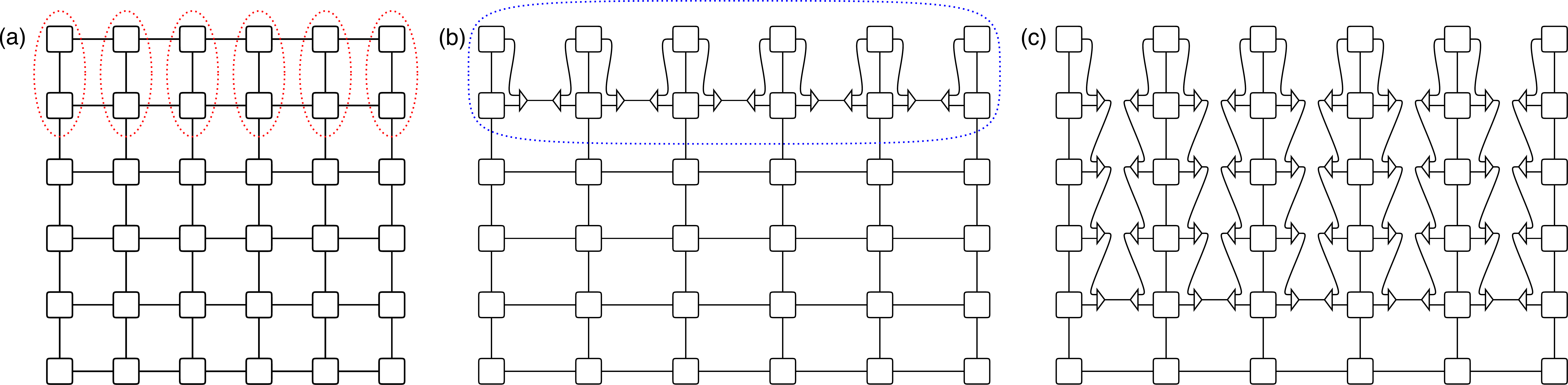}
\caption{(a) A two-dimensional partition function expressed as a tensor network. Boundary-MPS contraction proceeds by combining the top row of tensors with the next (red circles), and then (b) projecting the combined horizontal bonds to a subspace of maximal dimension $D$ (triangular tensors). Typically, the projectors are chosen to maximize the fidelity of resulting effective state of the upper-half of the system, outlined in blue. (c) After repetition, the end result is a tensor network that is contractible with cost linear in system size. \label{fig:2dpartition}}
\end{figure*}

The tensor network to contract is shown in \Fig{fig:2dpartition}, a closed tensor network diagram (with no open indices) representing a single number which in all cases I call the partition function $Z$. I denote the dimension of the bonds as $d$ and thus the rank-4 tensors in the bulk are $d\times d\times d\times d$ tensors. Contracting the tensor network and obtaining $Z$ exactly has a computational cost that scales exponentially with the lattice length $L$, as $O(d^L)$. To achieve this, the tensors in the first row could be multiplied with those in the second row, and then the third row, etc. At the $i$th step, the horizontal bond dimension has grown to $d^i$.

A more efficient, though approximate, way of contracting the tensor network is by truncating the horizontal bonds to some maximal dimension $D$. Specifically, in the renormalization schemes based on tensor networks, truncating a bond from dimension $R$ to $D$ is achieved by applying a pair of linear maps $W_L$ and $W_R$ from the $R$ dimensional space to a $D$ dimensional subspace --- these linear maps are simply $R\times D$ matrices and, typically, the pair form a projector $P = W_L W_R^{\dag}$ where $P^2 = P$ and $W_L$ is the quasi-inverse of $W_R^{\dag}$ (the projector is said to be orthogonal when $W_L = W_R^{\dag}$ are isometric matrices). The scheme is depicted in \Fig{fig:2dpartition}, where $W_L$ and $W_R$ are drawn as pairs of triangular tensors.

What remains is the determination of each of the $W_L$ and $W_R$. The boundary-MPS algorithm relies on minimizing the error of the boundary state made up of the all the outgoing vertical bonds from the upper region, outlined in blue in \Fig{fig:2dpartition}~(b). If the error is taken with respect to the 2-norm, then the optimal values of $W_L$ and $W_R$ can be determined directly. Take the state before truncation as $|\mathrm{MPS}\rangle$ and after $|\mathrm{MPS}^{\prime}\rangle$. We wish to solve the following minimization problem.
\begin{equation}
    \min_{W_L,W_R} \Bigl\| \, |\mathrm{MPS}\rangle - |\mathrm{MPS}^{\prime}\rangle \Bigr\|^2_2 \label{2norm}
\end{equation}

This can be solved most readily by using the Schmidt decomposition of the state,
\begin{equation}
    |\mathrm{MPS}\rangle = \sum_{i=1}^R S_i \; |L_i\rangle \otimes |R_i\rangle \label{schmidt}
\end{equation}
where the Schmidt vectors are orthonormal ($\langle L_i|L_j\rangle = \delta_{ij}$ and $\langle R_i|R_j\rangle = \delta_{ij}$) and the Schmidt coefficients can be chosen real and non-negative ($S_i \ge 0$). Any matrix-product state is readily put into this form, though to do so has a numerical cost scaling as the cube of the bond-dimension and, this procedure contributes the leading-order cost of the entire calculation. Details of how to perform this transformation can be found in the Appendix.

Once in this form, the optimal truncation is straightforward: we simply select the $D$ largest Schmidt coefficients $S_1,\dots,S_D$ and discard the remainder. This is achieved by a projector which is diagonal in the Schmidt basis. In the common case that the magnitude of the Schmidt coefficients decreases rapidly, the approximation can be extremely accurate for moderate values of $D$. In fact, for wide classes of physical systems it has been commonly observed that the error decays exponentially (or superpolynomially) with $D$, however this not true for critical systems, glassy systems, or systems of higher effective dimensionality. 

\subsection{Tensor Network Monte Carlo}

The boundary-MPS version of Tensor Network Monte Carlo proceeds exactly as above, except uses a non-deterministic algorithm to obtain $W_L$ and $W_R$ at each step. Although the subspace chosen by the projector will not always be ``optimal'' as above, this will allow us to, on average, take unbiased samples. 

They key to unbiased sampling to is to sum up all contributions to $Z$, big and small, in the limit of infinite samples. Given a set of possible samples $\mathcal{S}$, sample $s \in \mathcal{S}$ defines two $R\times D$ matrices $W_L^{(s)}$ and $W_R^{(s)}$ such that on average,
%\begin{equation}
%    \frac{1}{|\mathcal{S}|} \sum_{s\in\mathcal{S}} W_L^{(s)} W_R^{(s)\dag} =  \frac{1}{|\mathcal{S}|} \sum_{s\in\mathcal{S}} \raisebox{-0.2cm}{\includegraphics[width=1.1cm]{WWdag.pdf}} = \raisebox{-0.05cm}{\includegraphics[width=1.1cm]{Identity.pdf}} = I. \label{identity}
%\end{equation}
\begin{equation}
    \frac{1}{|\mathcal{S}|} \sum_{s\in\mathcal{S}} W_L^{(s)} W_R^{(s)\dag}  = I. \label{identity}
\end{equation}
Graphically, this relation is:
\begin{equation}
    \frac{1}{|\mathcal{S}|} \sum_{s\in\mathcal{S}} \raisebox{-0.2cm}{\includegraphics[width=1.1cm]{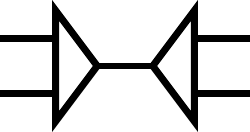}} = \raisebox{-0.05cm}{\includegraphics[width=1.1cm]{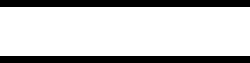}} \label{identity2}
\end{equation}
After sufficient sampling, the effect of the projectors is the identity operations --- in other words, there is no effect on the calculated partition function. On \emph{average}, \Fig{fig:2dpartition} (a), (b) and (c) will give the same result!

The choice of operators $W_L^{(s)}$ and $W_R^{(s)}$ that fulfils \Eqn{identity} is not unique. The goal now is to take a selection which minimizes the \emph{expectation value} of the error, so that the sample-to-sample variations are small. For instance, taking random projectors (with appropriate scaling) may be a solution of \Eqn{identity}, but it would be an extremely poor sampling scheme with unworkably large statistical variance, requiring a huge number of samples to achieve a reasonable accuracy.

To improve the statistical variance, importance sampling is used. In analogy to the choice of optimal projectors above, I use the Schmidt decomposition to fix the sampling basis and use the Schmidt coefficients $c_i$ to define the weights related to choosing $D$ of the $R$ possible basis vectors. Motivated by minimizing the expected 2-norm error [\Eqn{2norm}], I assign weight $c_i^2$ to the the probability of selecting the $i$th vector. The weight of simultaneously choosing vectors $i_1, i_2, \dots, i_D$ is given by their product,
\begin{equation}
    w(i_1,i_2,\dots,i_D) = \prod_{j=1}^D S_{i_j}^2. \label{weight}
\end{equation}
The probability of a certain selection $\mathbf{i} = [i_1,\dots, i_D]$ is the normalized form of these weights,
\begin{equation}
    p(\mathbf{i}) = w(\mathbf{i}) / \sum_{\mathbf{i}} w(\mathbf{i}). \label{probability}
\end{equation}
This probability distribution is known in the literature as ``Fischer's multivariate, non-central hypergeometric distribution'' where each element can be selected at most once. It is also related to the canonical Fermi-Dirac distribution (i.e. with fixed particle number, $D$). How to sample efficiently and directly from this distribution is not immediately obvious, but it is possible to write the probability distribution itself as a simple matrix-product state from which perfect sampling can be performed~\cite{Ferris2012,Stoudenmire2010}. See the Appendix for details and the freely available sampling code. 

The advantage of this distribution is that the Schmidt coefficients with largest weights are almost-always selected, while the tails of the distribution are only selected occasionally. This gives us most of the accuracy of the simple-truncation approach, while sampling removes the bias inherited from deterministically throwing away part of the boundary state. However, to ensure the requirements of \Eqn{identity} are met, we must scale each component $k$ by the inverse of its appearance rate $1/r(k)$, where
\begin{equation}
    r(k) = \sum_{k\in\mathbf{i}} p(\mathbf{i}) \label{rate}
\end{equation}
The values of $r(k)$ are readily extracted from the matrix-product form of Fischer's hypergeometric distribution. The resulting $W_L$ and $W_R$ do not form a projector; nonetheless, they are expected to lead to single-sample accuracies that are comparable to the standard truncation approach. The entire sampling cost scales linearly in the product $DR$ and is not a leading-order cost of the simulation.

It is essential that the projectors at different points in space are chosen independently. Because of this, TNMC is limited to simulations of finite systems, which is one drawback compared to variational tensor network approaches which may directly address the thermodynamic limit. 

At the end of the calculation, an estimate of the partition function $Z_{\mathbf{s}}$ is produced. To get an estimate of $Z$ and of its statistical uncertainty, multiple samples should be taken to determine the mean and standard error. 

\section{Numerical results}

\subsection{Statistical error versus bond dimension}

\begin{figure}[t]
\centering
\includegraphics[width=0.85\columnwidth]{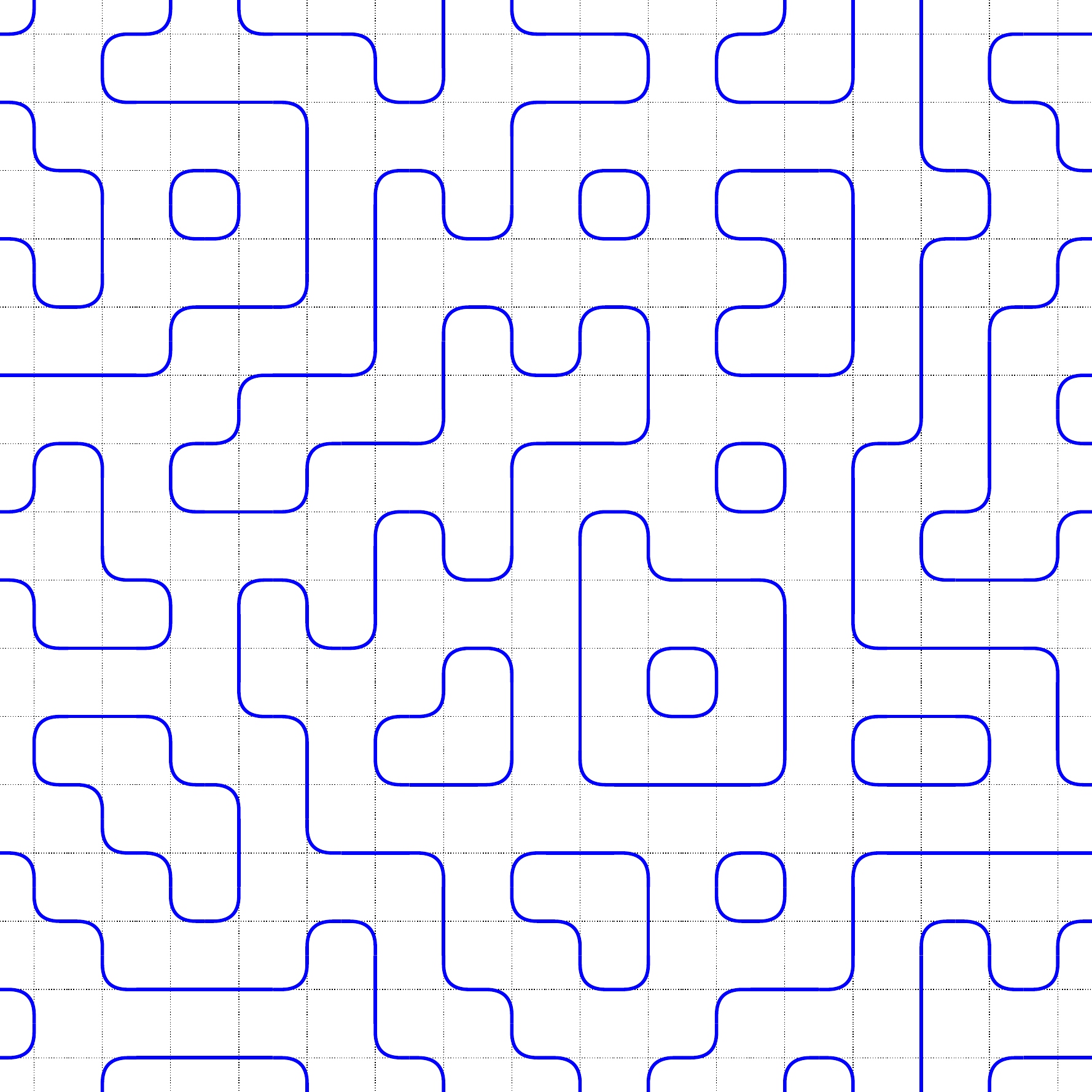}
\caption{One of the approximately $7.05\times10^{56}$ loop configurations on a $16\times 16$ lattice with open boundary conditions. \label{fig:sixvertex16x16}}
\end{figure}

\paragraph{The model:} We first study the partition function of a simple fully-packed loop model on a 2D lattice, which represents the counting of all possible closed-loop configurations where each vertex must have a single line entering and exiting, represented locally by these six configurations:
{\center \includegraphics[width=\columnwidth]{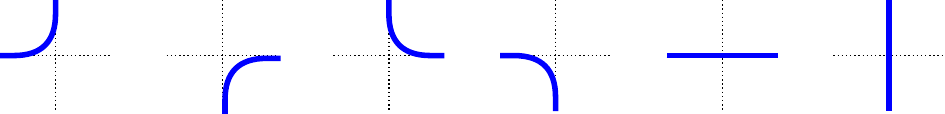}\\}
An example configuration on an open $16 \times 16$ lattice is depicted in Fig.~\ref{fig:sixvertex16x16}.
The model has one one-to-one correspondence to the six-vertex model as well as the counting of alternating sign matrices, and is integrable (solvable) via Bethe ansatz techniques. The number of permissible loop configurations can be deduced by creating a ``partition function'', to which the value~1 is added for every allowable configuration. The partition function can be expressed as a two-dimensional network of connected tensors. At each vertex, a $2\!\times\!2\!\times\!2\!\times\!2$ tensor $T$ is placed and connected to it's nearest neighbours. The value $T_{ijkl} = 0$ unless it corresponds to one of the six configurations depicted above, that is, $T_{1100} = T_{1001} = T_{0110} = T_{0011} = T_{1010} = T_{0101} = 1$. The number of allowable configurations grows very fast in the lattice size. This calculation is selected because of its simplicity and accessibility to a wide audience, and because its numerical evaluation is moderately challenging.

\begin{figure}[t]
\centering
\includegraphics[width=0.95\columnwidth]{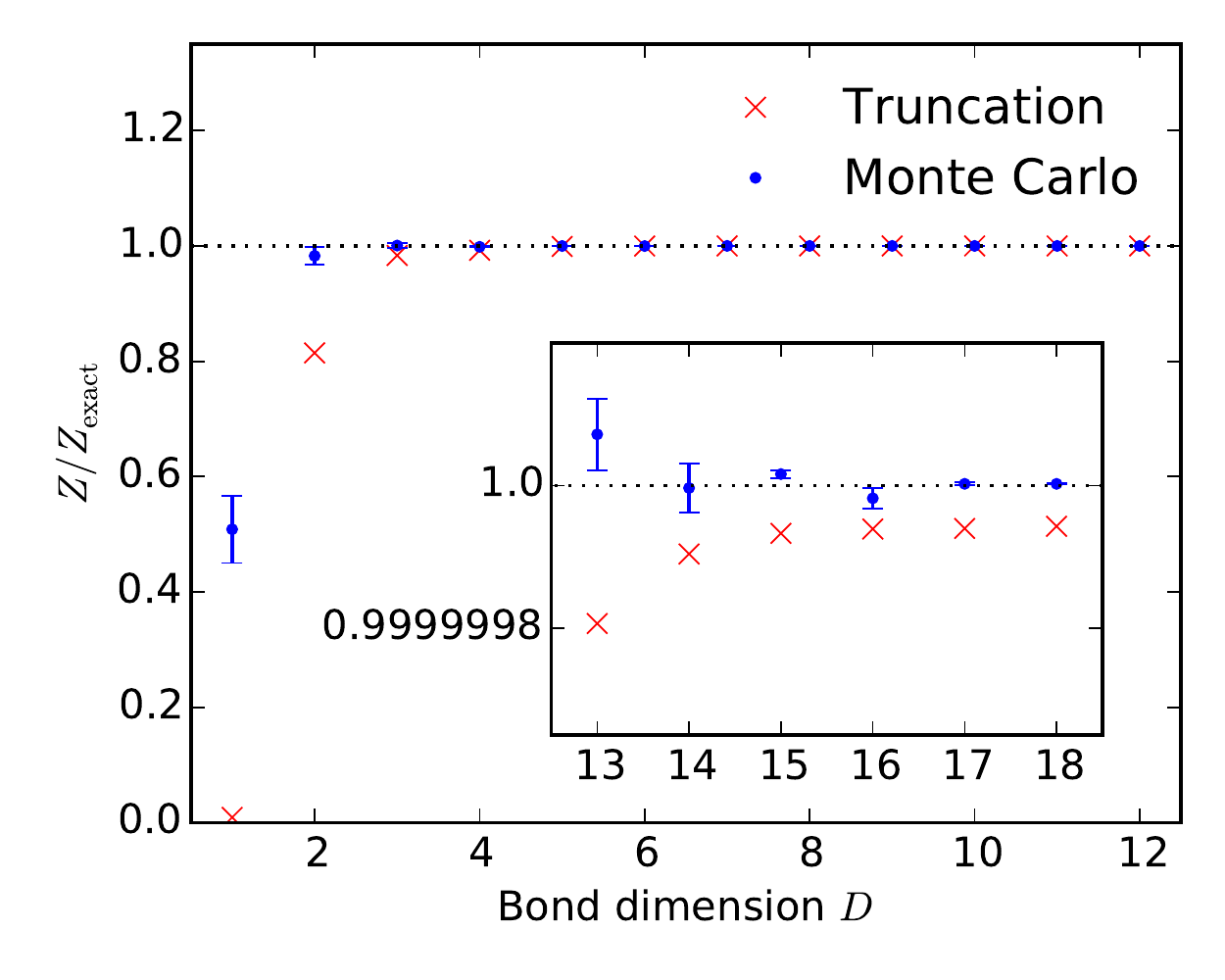}
\caption{Calculations of the partition function of the six-vertex model on a 16x16 lattice (where $Z_{\mathrm{exact}} \approx 7.05\times10^{56}$), using standard MPS truncation and $10^4$ tensor network Monte Carlo samples with the given bond dimension $D$. The TNMC results are within one or two standard deviations to the exact results, independent of bond dimension (the $D=1$ result suffers from broad tails on the distribution, due to our perfect sampling approach). In both cases, the errors are reduced with increasing bond dimension. \label{fig:sixvertex1}}
\end{figure}

\paragraph{Results:} In \Fig{fig:sixvertex1}, I plot the results for the calculation of the partition function (or number of loop configurations), compared to the quasi-exact result extracted from a boundary-MPS with large bond dimension. Here I have studied a modest $16
\times 16$ lattice with open boundary conditions, which allows for appoximately $7.05 \times 10^{56}$ configurations. While boundary-MPS calculations using the standard truncation scheme continually underestimates the partition function, the Tensor Network Monte Carlo estimates fluctuate about the exact result (in this case, determined quasi-exactly by an MPS calculation with large bond dimension) independently of bond dimension.

\begin{figure}[t]
\includegraphics[width=0.95\columnwidth]{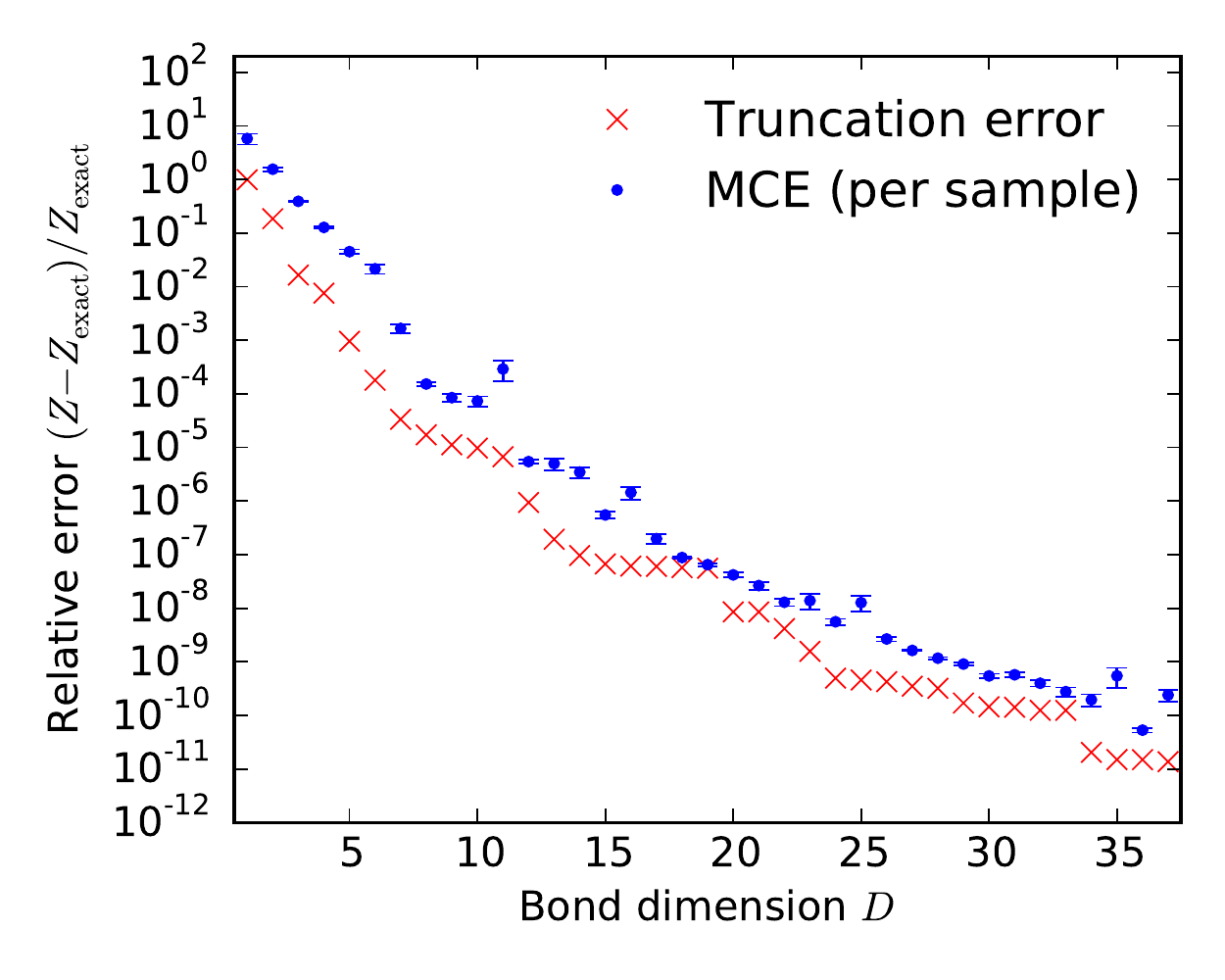}
\caption{Statistical and truncation error for the partition function of the six-vertex model, as in \Fig{fig:sixvertex1}. Unlike standard Monte Carlo, a single tensor network Monte Carlo sample may account for the vast majority of the partition function, therefore reducing the statistical fluctuations. The standard deviation between TNMC samples parallels the behaviour of the traditional tensor network contraction, the error decaying extremely rapidly with respect to bond dimension. \label{fig:sixvertex2}}
\end{figure}

What does change with bond-dimension is the accuracy or precision of both simulations, illustrated in Fig.~\ref{fig:sixvertex2}. The standard tensor network technique converges to the exact result extremely rapidly as the bond dimension is increased; in fact, for $D \ge 40$ we have already saturated numerical precision (larger $D$ may be required for bigger systems or periodic boundary conditions). What we see clearly from Fig.~\ref{fig:sixvertex2} is that the TNMC method inherits similar behaviour --- as the bond dimension increases, the sample-to-sample variations decrease just as rapidly as the standard tensor network approach (although the standard deviation may be a small constant larger in magnitude than the variational error, which is the cost paid to obtain an unbiased simulation). This is an enormous improvement over standard, configuration-based Monte Carlo, where the sample-to-sample variance is fixed by the problem or physical system you are trying to solve.

The TNMC method now gives the user two avenues to decrease simulation error --- increasing either bond dimension, or the number of samples. While performing $N$ samples decreases the overall error by a factor of $1/\sqrt{N}$, increasing $D$ may result in a much more significant improvement (often super-polynomial). Nonetheless, there are many situations where there are insufficient computational resources to utilize a large enough bond-dimension for accurate, trustworthy and unbiased variational results. In this case, the TNMC may be useful to remove the bias of standard tensor network simulations, which I investigated next.

\subsection{Removal of variational bias}

\paragraph{The model: } Next we study the classical Ising model on a square lattice at zero magnetic field. At each vertex $i$ of the lattice, a spin $s_i$ can take values $\pm1$. The Hamiltonian is a sum of two-body contributions, where nearest neighbours on the lattice contribute energy $-1$ if they align and $+1$ if they anti-align. Formally, the Hamiltonian is
\begin{equation}
    H = -J \sum_{<i,j>} s_i s_j, \label{isinghamiltonian}
\end{equation}
where $<\!\!i,j\!\!>$ denote nearest neighbour pairs, $i$ and $j$. Without loss of generality, I take the coupling constant $J = 1$ and Boltzmann's constant $k_B = 1$. At temperature $T$, the probability of a configuration $\mathbf{s} = [s_1, s_2, \dots]$ is $p(\mathbf{s}) =\exp(-H(\mathbf{s})/T) / Z$, where $Z$ is the partition function is the normalization constant of all the individual weights:
\begin{equation}
    Z = \sum_{\mathbf{s}} \exp(-H/T) = \sum_{\mathbf{s}} \prod_{<i,j>} \exp(-s_i s_j / T).
\end{equation}
This result can be expressed as a two-dimensional tensor network in the form studied above, with rank-four tensors on the vertices of a square lattice (see the Appendix for further details).

This well-known model exhibits a phase transition from a disordered phase (randomly aligned spins) to ferromagnetic phase (all spins aligned in the same direction, on average) at a critical temperature $T_c\approx 2.278$. At this point $Z$ becomes non-analytic in the thermodynamic limit, and the symmetry is broken for $T<T_c$ (the system may either be majority spin-up or majority spin-down). Here, I will study finite-sized $L\times L$ systems (L odd) with \emph{closed} boundary conditions (the boundary consisting of all up spins), resulting in finite-size effects that smooth out the phase transition. The spontaneous magnetization $M$, as measured by the average magnetization $M =\overline{s_c}$ of the central spin $c$, is a measure of the order parameter. I have chosen this configuration because of its convenience with respect to the TNMC using boundary MPS, although periodic boundaries would result in smaller finite-size effects. The central spin is measured by contracting two boundary-MPSs, one from the top boundary and another independently from the bottom boundary. The final contraction involves the central row surrounded by the two boundary MPS and \emph{does not involve renormalization of the spin being measured}, allowing us to easily measure both $Z$ and the function $Z_M$
\begin{equation}
    Z_M = \sum_{\mathbf{s}} s_c \exp{(-H/T)}
\end{equation} 
which gives the spontaneous magnetization $M = Z_M / Z$. In the Monte Carlo simulation, the numerator and denominator must be averaged \emph{before} division, and because of this the jackknife technique is used here to estimate the statistical error. It should be noted that wild fluctuations were observed when attempting to renormalize the measured spin, by including it in rows which are truncated in a simple top-to-bottom contraction, as the strong correlation between $Z_M$ and $Z$ is lost (resulting in single-sample estimates of $M$ that may be well outside the ``physical'' region $0 \le M \le 1$, even though averages after many samples may be correct).

\begin{figure}[t]
\includegraphics[width=0.95\columnwidth]{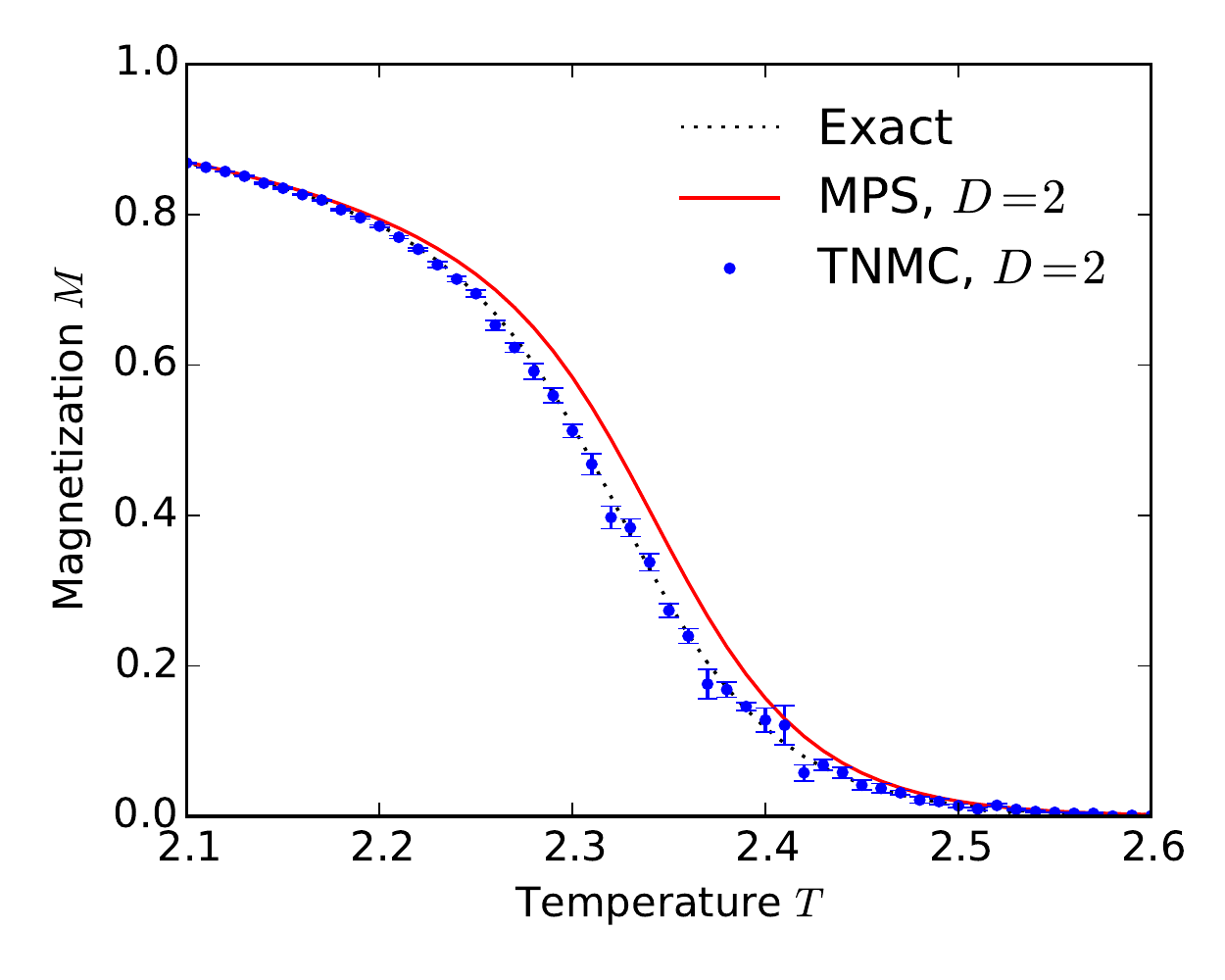}
\caption{The central magnetization of the classical 2D Ising model on a $63\times63$ lattice with closed boundary conditions as a function of temperature. The standard MPS method with bond dimension $D=2$ overestimates the magnetization and thus transition temperature due to the bias introduced by truncation. The TNMC with same bond dimension and $10^3$ samples gives an unbiased result, however the statistical fluctuations are greatest in the same region where the biased technique struggles. \label{fig:ising1}}
\end{figure}

\paragraph{Results: } In Fig.~\ref{fig:ising1}, I plot the spontaneous magnetization of a $63\times63$, comparing boundary-MPS results with standard truncation and TNMC using bond dimension $D=2$ to quasi-exact results (obtained with large bond dimension). The result is clear --- the variational MPS technique overestimates both the spontaneous magnetization and the temperature of the phase transition, however the TNMC results fluctuate about the exact curve as expected. The variational result prefers the ``simpler'' ferromagnetic phase since it is somehow ``easier'' to capture with a small bond-dimension (here, the symmetry is explicitly broken by the boundary condition). This type of bias is a well-know problem with all variational techniques. On the other hand TNMC result allows for spatial fluctuations much like standard Monte Carlo, and has no preference for one phase over the other. The tensor network bias has been completely removed.

\section{Discussion}

\subsection{Markov chain sampling}

The method above has relied on so-called ``perfect sampling'', where each sample is drawn independently from the last. This has many benefits, in that it is simple to parallelize and the independence of samples makes statistical analysis straightforward.

However, perfect sampling may also have some important limitations. The distribution of the estimates $Z_\mathbf{s}$ is well-controlled for sufficiently large bond-dimension, but for small bond-dimensions and very large systems the statistics may be very broad and suffer from long tails. In this case, very unlikely samples may make very significant contributions to the partition function, and worse, the measured variance may not correspond to the true error. The observation made during the simulations presented here is that perfect sampling is effective whenever the bond-dimension is large enough to capture a large fraction of the partition function (even for the variational method). This becomes more problematic for larger systems because $Z$ increases exponentially, not linearly, with system size, and a constant fraction of $Z$ is necessarily lost at each truncation step. TNMC techniques at least have the possibility to overcome this limitation because their accuracy may also improve exponentially in the bond-dimension. On the other hand, standard Monte Carlo techniques almost \emph{never} utilize perfect sampling because of the inherent difficulty in choosing relevant samples according to the partition function.

The alternative to perfect sampling is Markov chain sampling, where the configuration is updated iteratively, and is preferred because it rapidly stabilizes to ``likely'' configurations according to the partition function. The Metropolis algorithm is perhaps the most well-known Markov chain sampling technique, where a new configuration $\mathbf{s}^\prime$ is accepted with a probability that depends on the ratio of $Z_{\mathbf{s}^\prime}/Z_{\mathbf{s}} $. 

A similar technique could be applied to TNMC, but must first be adapted to the multi-sampling that takes place at each projection. The tensor-network algorithm would work in a similar fashion to the so-called ``second-renormalization"' group (SRG) where the projectors are optimized iteratively. However, now the selected subspace would be chosen according to the \emph{product} of their individual weights, similar to Eqns.~(\ref{weight},\ref{probability}), and then individually rescaled according to their rates [\Eqn{rate}]. One expects that this would improve both the quality of each sample (i.e. how much of the partition function is accounted for in a single sample, similar to how the SRG improves over TRG) and the variance of observables, at the expense of some autocorrelation between samples.

\subsection{Other tensor network renormalization schemes}

The TNMC can be applied in a straightforward way to a variety of geometries using existing renormalization schemes. In the corner transfer matrix (CTM) and the various tensor renormalization group (TRG) techniques, the renormalization scheme proceeds by iteratively enlarging the course-grained blocks and truncating the bonds of the resulting tensor network. Instead of choosing the subspace that maximizes the fidelity of the representation of the blocks, one could choose them randomly according to their individual importance using the core of the TNMC algorithm presented here.

One particularly exciting prospect would be the application to 3D partition functions using 3D TRG algorithms. The ``higher-order'' TRG (HOTRG) method in particular has been used to obtain very accurate results. Despite this, HOTRG is known to \emph{not} account for all short-range correlations according to the `area law' in three or higher dimensions~\cite{Ferris2013}, in analogy to the way that DMRG and tree tensor networks are unable to account for all the correlations of very large 2D quantum systems.

One way of avoiding any doubts surrounding 3D HOTRG raised by this problem is by using TNMC to remove any potential bias, while the renormalization scheme itself is used to average a significant part of the partition function --- even if it is not able to be account for an absolute majority of $Z$ with a single sample, statistical variations of observables will be \emph{much} smaller than in standard, configuration-based Monte Carlo, especially those observables measured in the well-renormalized regions of space. This development would be exciting because it would open the possibility of precise and unbiased calculations of challenging 3D classical, and 2D quantum, systems.

\subsection{Quantum systems and real-time evolution}

Any $d$-dimensional quantum statistical problem can be expressed, at least approximately, as a $(d+1)$-dimensional tensor network called the `path integral' via a decomposition of the imaginary time evolution, $\exp(-\hat{H}/T)$. Performing (quantum) Monte Carlo by sampling from the resulting tensor network is known as Path Integral Monte Carlo (PIMC). PIMC has been wildly successful in many cases, but suffers from the sign problem in higher-dimensional frustrated or fermionic systems.

TNMC can improve upon PIMC in two ways. Firstly, even in the absence of a sign problem, TNMC will display much smaller statistical variations than traditional PIMC, and may be a much more economical method of generating results of a certain precision. Secondly, the HOTRG has been shown to mitigate the sign problem given a sufficiently large bond-dimension~\cite{Denbleyker2014}, allowing the summation of both positive and negative contributions to the partition function to obtain the correct result. Similarly, each TNMC sample generated via 3D HOTRG would also sum over many of the positive and negative contributions, potentially eliminating the sign problem entirely if the bond-dimension is sufficient. The Monte Carlo aspect allows us to parallelize work over many computers and, by averaging many samples, obtain a higher precision than is possible with the variational HOTRG method. The possibility of creating a sign-problem resistant Monte Carlo to attack challenging 2D quantum problems is certainly enticing, and is an area of intense interest for the author.

The extra `imaginary-time' dimension could be replaced with real-time evolution for simulating unitary quantum dynamics, the stochastic dynamics of classical systems, or some combination of both. It is expected that our ability to perform unitary quantum dynamics will be limited --- ultimately by the sign problem and because correlations in time do not decay under unitary dynamics. The dynamics of open systems is interesting both theoretically and because of the relevance to modern physics experiments, and given that correlations \emph{do} decay in time as the system relaxes to the steady state, the TNMC should work well.

One could speculate further, and foresee continuous-time Tensor Network Monte Carlo techniques such as projector Monte Carlo, stochastic series expansion or diagrammatic Monte Carlo, or anticipate a connection between real-time evolution of MPS and the Monte Carlo trajectories of simpler variational wavefunctions, for instance, by the Positive-P simulation method that uses the bosonic coherent state basis~\cite{Drummond1980}.

\section{Conclusion}

I have introduced an entire new class of calculation under the umbrella of Tensor Network Monte Carlo. TNMC is capable of producing significantly less sample-to-sample variance than standard Monte Carlo through what is essentially an aggressive multi-sampling technique that can account for the majority of the partition function \emph{in a single sample}. Although this approach is based on methods for renormalizing tensor networks, TNMC exhibits none of the variational bias of those techniques. 

The core of the algorithm is modular by nature and can be readily retrofitted to other tensor network renormalization schemes beyond the boundary-MPS calculations presented here. The most obvious application would be to Tensor Renormalization Group calculations for 2D and 3D partition functions. The method naturally applies to quantum systems via the path integral formulation, and can potentially by used to approach a huge array of problems in physics and beyond. It is natural to attempt the simulation of strongly-correlated, frustrated or fermionic quantum problems in 2D --- and ask if we can achieve the `holy grail' of quantum Monte Carlo by performing calculations which are normally plagued by the sign problem.

\subsection{Acknowledgement}

I would like to thank Luca Tagliacozzo for stimulating discussions. This work was supported by Marie Curie fellowship SQSNP 622939 FP7-MC-IIF, TOQATA (Spanish grant PHY008-00784) and the EU IP SIQS.

\bibliography{../bib/andy}

\begin{thebibliography}{99}

% Monte Carlo coined
\bibitem{Metropolis1949}
N. Metropolis and S. Ulam, % The Monte Carlo method
Journal of the American Statistical Association {\bf 44}, 335 (1949).

% Metropolis algorithm
\bibitem{Metropolis1953}
N. Metropolis, A.W. Rosenbluth, M.N. Rosenbluth, A.H. Teller, and E. Teller, % "Equations of State Calculations by Fast Computing Machines". 
Journal of Chemical Physics \textbf{21}, 1087 (1953).

% SW first cluster updates
\bibitem{Swendsen1987}
R. H. Swendsen and J.-S. Wang, % Nonuniversal critical dynamics in Monte Carlo simulations. 
Phys. Rev. Lett. \textbf{58}, 86 (1987).

% Wolff algorithm cluster update
\bibitem{Wolff1989}
Ulli Wolff, % "Collective Monte Carlo Updating for Spin Systems", 
Phys. Rev. Lett. \textbf{62}, 361 (1989).

% First loop-update MC
\bibitem{Evertz1993}
H. G. Evertz, G. Lana, and M. Marcu, Phys. Rev. Lett. {\bf 70}, 875 (1993).
% And review
\bibitem{Evertz2003}
H. G. Evertz, Adv. Phys. {\bf 52}, 1 (2003).

% Worm algorithm
\bibitem{Prokofev1998}
N.V Prokof'ev, B.V Svistunov, and I.S Tupitsyn, % ?Worm? algorithm in quantum Monte Carlo simulations
Physics Letters A {\bf 238}, 253 (1998).

% Multisampling
\bibitem{Sankowski2003}
Piotr Sankowski, in \emph{Algorithms --- ESA 2003}, edited by G. De Battista and U. Zwick (Springer Berlin Heidelberg, 2003), pp. 740--751.

%DMRG
\bibitem{White1992}
S.R. White, Phys. Rev. Lett. {\bf 69}, 2863 (1992).

% DMRG = MPS
\bibitem{Ostlund1995}
S. {\"O}stlund and S. Rommer, %Thermodynamic Limit of Density Matrix Renormalization
Phys. Rev. Lett. {\bf 75}, 3537 (1995).

% Boundary MPS for classical problems
\bibitem{Orus2008}
R. Or{\'u}s and G. Vidal, %Infinite time-evolving block decimation algorithm beyond unitary evolution, 
Phys. Rev. B {\bf 78}, 155117 (2008).

% CTM = DMRG
\bibitem{Nishino1997}
T. Nishino and K. Okunishi, %Corner Transfer Matrix Algorithm for Classical Renormalization Group
J. Phys. Soc. Jpn. {\bf 66}, 3040 (1997).

% TTN
\bibitem{Shi2006}
Y. Shi, L. Duan, and G. Vidal, Phys. Rev. A {\bf 74}, 022320 (2006).
\bibitem{Tagliacozzo2009}
L. Tagliacozzo, G. Evenbly, and G. Vidal, Phys. Rev. B {\bf 80}, 235127 (2009).

% PEPS
\bibitem{Verstraete2004}
F. Verstraete and J. I. Cirac, eprint arXiv:cond-mat/0407066 (2004).
\bibitem{Jordan2008}
J. Jordan, R. Orus, G. Vidal, F. Verstraete, and J. I. Cirac, Phys. Rev. Lett. {\bf 101}, 250602 (2008).
\bibitem{Gu2008}
Z.-C. Gu, M. Levin, and X.-G. Wen, Phys. Rev. B {\bf 78}, 205116 (2008).

% MERA
\bibitem{Vidal2007b} 
G. Vidal, Phys. Rev. Lett. {\bf 99}, 220405 (2007).
\bibitem{Vidal2008}
G. Vidal, Phys. Rev. Lett. {\bf 101}, 110501 (2008).

% Branching MERA
\bibitem{Evenbly2014b}
G. Evenbly and G. Vidal, %A class of highly entangled many-body states that can be efficiently simulated
Phys. Rev. Lett. {\bf 112}, 240502 (2014).

% Spectral TN
\bibitem{Ferris2014}
A.J. Ferris, %Fourier transform of fermionic systems and the spectral tensor network
Phys. Rev. Lett. {\bf 113}, 010401 (2014).

% 2D results
\bibitem{Yan2010}
Simeng Yan, David A. Huse, and Steven R. White, %Spin Liquid Ground State of the S=1/2 Kagome Heisenberg Model
Science {\bf 332}, 1173 (2011).
\bibitem{Corboz2014}
P. Corboz and F. Mila, %Crystals of bound states in the magnetization plateaus of the Shastry-Sutherland model
Phys. Rev. Lett. {\bf 112}, 147203 (2014).
\bibitem{Corboz2014b}
Philippe Corboz, T. M. Rice, and Matthias Troyer, %Competing states in the t-J model: uniform d-wave state versus stripe state
Phys. Rev. Lett. 113, 046402 (2014).

% TRG
\bibitem{Levin2006}
M. Levin and C.P. Nave, %Tensor renormalization group approach to 2D classical lattice models
Phys. Rev. Lett. {\bf 99}, 120601 (2007).

% HOTRG
\bibitem{Xie2012}
Z. Y. Xie, J. Chen, M. P. Qin, J. W. Zhu, L. P. Yang, and T. Xiang, %Coarse-graining renormalization by higher-order singular value decomposition
Phys. Rev. B {\bf 86}, 045139 (2012).

% TNR
\bibitem{Evenbly2014}
G. Evenbly and G. Vidal, %Tensor Network Renormalization
e-print arXiv:1412.0732 (2014).

% Fixed-node Projector-Monte Carlo with MPS guiding WFs
\bibitem{Wouters2014}
S. Wouters, B. Verstichel, D. Van Neck, and G.K.-L. Chan, %Projector quantum Monte Carlo with matrix product states
Phys. Rev. B 90, 045104 (2014)
\bibitem{Sikora2015}
O. Sikora, H.-W. Chang, C.-P. Chou, F. Pollmann, and Y.-J. Kao, %Variational Monte Carlo simulations using tensor-product projected states
Phys. Rev. B {\bf 91}, 165113 (2015).

% Variational Monte Carlo for TNs
\bibitem{Sandvik2007}
A. W. Sandvik and G. Vidal, %Variational quantum Monte Carlo simulations with tensor-network states
Phys. Rev. Lett. {\bf 99}, 220602 (2007).
\bibitem{Wang2011}, %Monte Carlo simulation with Tensor Network States
Ling Wang, Iztok Pizorn, and Frank Verstraete,
Phys. Rev. B {\bf 83}, 134421 (2011).
\bibitem{Ferris2012}
A.J. Ferris and G. Vidal, %Perfect Sampling with Unitary Tensor Networks
Phys. Rev. B {\bf 85}, 165146 (2012).
\bibitem{Ferris2012b}
A.J. Ferris and G. Vidal, %Variational Monte Carlo with the Multi-Scale Entanglement Renormalization Ansatz
Phys. Rev. B {\bf 85}, 165147 (2012).
\bibitem{Iblisdir2014}
S. Iblisdir, %Simulated annealing for tensor network states
New J. Phys. {\bf 16} 103022 (2014).

% METTS details + perfect sampling of MPS is D^2
\bibitem{Stoudenmire2010}
E.M. Stoudenmire and S.R. White, % Minimally Entangled Typical Thermal State Algorithms
New J. Phys. {\bf 12}, 055026 (2010).

% Area law and 3D TRG
\bibitem{Ferris2013}
A.J. Ferris, %The area law and real-space renormalization
Phys. Rev. B {\bf 87}, 125139 (2013).

% TRG is sign-problem resistant
\bibitem{Denbleyker2014}
A. Denbleyker, Y. Liu, Y. Meurice, M. P. Qin, T. Xiang, Z.Y. Xie, J.F. Yu, and H. Zou, %Controlling sign problems in spin models using tensor renormalization
Phys. Rev. D {\bf 89}, 016008 (2014).

\bibitem{Drummond1980}
P. D. Drummond and C. W. Gardiner, %Generalized P-representations in quantum optics. 
J. Phys. A: Math. Gen. {\bf 17}, 2353 (1980).

\end{thebibliography}

\section{Appendix}

\subsection{Sampling basis}

The basis selected to perform sampling is selected in the same fashion as the standard truncation approach, via putting the boundary state in Schmidt form, which we rewrite.
\begin{equation}
    |\mathrm{MPS}\rangle = \sum_{i=1}^R c_i \; |L_i\rangle \otimes |R_i\rangle
\end{equation}
The left Schmidt vectors $|L_i\rangle$ and also the right Schmidt vectors form orthonormal bases, $\langle L_i | L_j \rangle = \delta_{ij} = \langle R_i | R_j \rangle$.

Getting the MPS into this form is relatively straightforward. Beginning on the left half of the chain, one uses the $QR$-decomposition on the first tensor, $A_1$, such that $A_1 = Q_1 R_1$:
\begin{equation}
    \includegraphics[height=5ex]{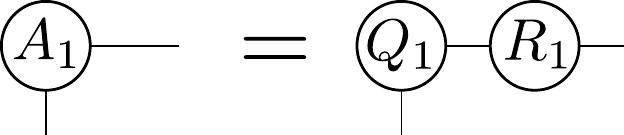}
\end{equation}
where $R$ is an upper-triangular (possibly rectangular) matrix while $Q_1$ is an isometric operator $Q_1 Q_1^{\dag} = I$: 
\begin{equation}
    \includegraphics[height=10ex]{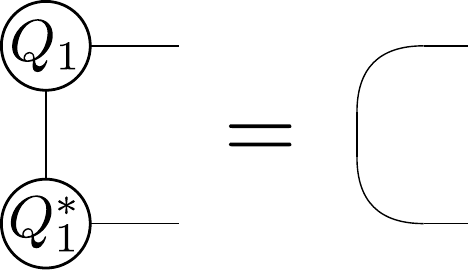}
\end{equation}
For the same purpose one could utilize the singular-value decomposition (SVD) but the $QR$-decomposition is numerically less demanding. At the next step we perform the $QR$-decomposition on $R_1 A_2$:
\begin{equation}
    \includegraphics[height=5ex]{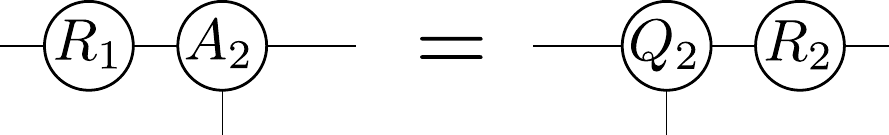}
\end{equation}
where:
\begin{equation}
    \includegraphics[height=10ex]{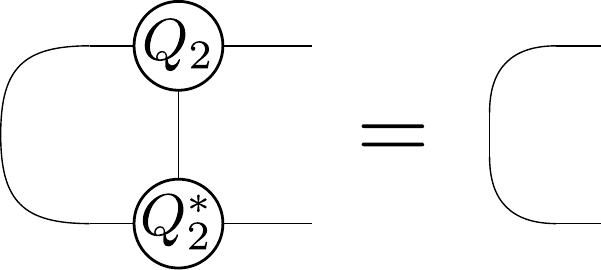}
\end{equation}
and so-on, until we reach the desired bond $i$ with generated $R_i$. We repeat the procedure from the right, using the related `$LQ$-decomposition' where $L$ is lower-triangular and $Q$ is again isometric (the $LQ$-decomposition of $M$ is easily related to the $QR$-decomposition of $M^T$ via the transpose operation). The process is repeated until we find $L_{i+1}$.

We are left with an MPS in a form with all the MPS tensors in either left- or right-isometric form, depending on if they are to the left or right of the bond we wish to truncate. On this bond, we have an additional matrix $R_i L_{i+1}$, which we can decompose according the the SVD:
\begin{equation}
    R_i L_{i+1} = U S V,
\end{equation}
where $U$ and $V$ are unitary and $S$ is diagonal and non-negative. The matrix $U$ can be absorbed by the tensor to the left, without changing its isometric nature, and similarly $V$ can be absorbed to the right. We are then left with the following form of the MPS, written for 6 sites,
\begin{equation}
    \includegraphics[height=5ex]{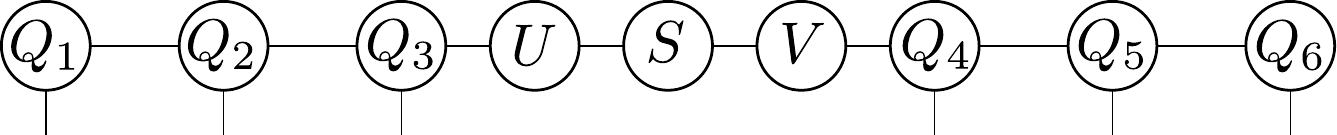}
\end{equation}
where the left half forms an orthonormal basis due to the isometric nature of the tensors
\begin{equation}
    \includegraphics[height=10ex]{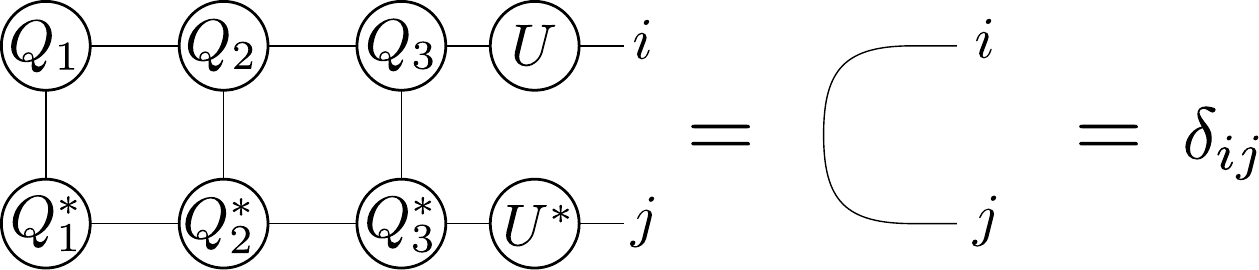}
\end{equation}
and similarly for the right
\begin{equation}
    \includegraphics[height=10ex]{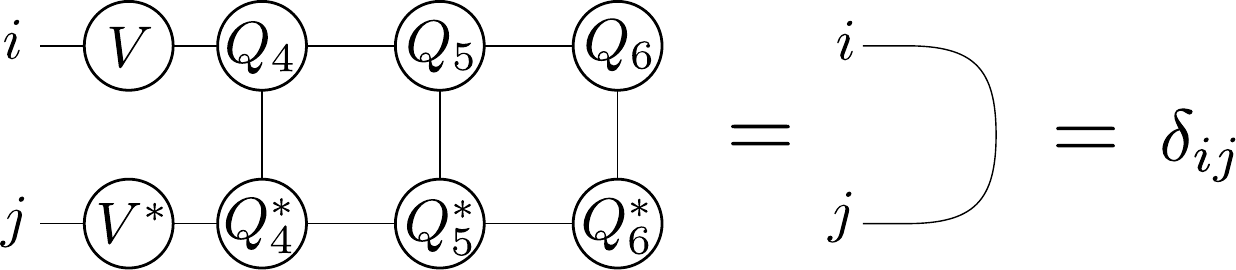}
\end{equation}

Finally, we truncate the bond by choosing to keep only a subset of the singular values (Schmidt vectors). In standard MPS truncation, we choose the largest $D$ of the $R$ singular values, resulting in an error according to the 2-norm as
\begin{equation}
     \Bigl\| \, |\mathrm{MPS}\rangle - |\mathrm{MPS}\rangle \Bigr\|^2_2  = \sum_{i = D^{\prime}+1}^D S_i^2
\end{equation}
where the $S_i$ are in descending order. Typically, we observe the distribution of singular values decays quickly --- often exponentially --- meaning the approximation error is small. In the TNMC, we use this basis since it results in a small \emph{expectation value} of the 2-norm error (i.e. the error averaged over many samples). It should be noted that using this basis is not a necessary requirement of TNMC --- strictly speaking, any basis, even overcomplete bases, can be used along with any sampling probabilities so long as \Eqn{identity} is reproduced.

After the sample is chosen, we move onto another bond and repeat. In practice, all bonds can be truncated with a single round of $QR$-decompositions to the right followed by the SVD sweeping to the left, truncating with independent Monte Carlo samples at each stage.

\subsection{Sampling algorithm}

In a single truncation step of TNMC, a sample refers to the simultaneous selection of some subset of size $D$ drawn from $R$ possibilities. Let us refer to that subset as the vector $\mathbf{i} = [i_1,i_2,\dots,i_{D}]$ where each element labels a selected basis vector. The unnormalized weight of the sample is
\begin{equation}
    w(i_1,i_2,\dots,i_D) = \prod_{j=1}^{D} S_{i_j}^2, \label{sampling_appendix}
\end{equation}
while the normalized probability is
\begin{equation}
    p(\mathbf{i}) = w(\mathbf{i}) / \sum_{\mathbf{i}} w(\mathbf{i}).
\end{equation}

This selection can be seen as a kind of \emph{multisampling}, where we imagine independently choosing $D$ samples according to their relative weights, $S^2_i$. However, we must additionally apply the constraint that no index is selected more than once. Such a constraint can easily be implemented as a matrix-product operator (MPO).

We begin with a product distribution, where each index $i$ contributes weight $S_i^2$ to the product if selected, and $1$ otherwise. This is simply represented by the vector $\mathbf{w^{(i)}} = [1, S_i^2]$, where we start indexing at zero ($w^{(i)}_0 = 1$, $w^{(i)}_1 = S_i^2$). To this we then apply the constraint MPO, pictured as
\begin{equation}
    \includegraphics[height=10ex]{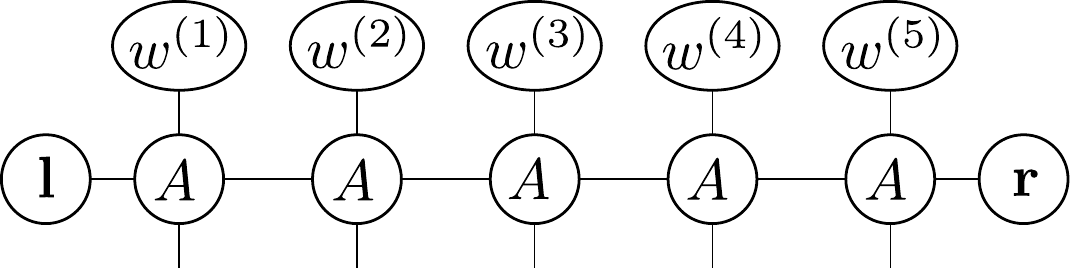}
\end{equation}
The tensors in the constraint MPO are simply given by $A_{s,s^\prime, i, j} = \delta_{s,s^{\prime}} \delta_{i+s,j}$, where $A$ is labelled:
\begin{equation}
    \includegraphics[height=10ex]{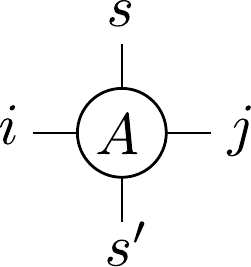}
\end{equation}
At the ends, we cap the MPO with vectors $\mathbf{l} = [1, 0, 0, \dots]$ to select that we ``begin'' with zero selections, and $\mathbf{r} = [0,\dots,0,1]$ denoting that we ``end'' with $D$ selections.

The product of the MPO and the product state results in an MPS, which can be sampled from easily with total cost $D R$ by taking advantage of the fact that $A$ is very sparse. To do this, one sequentially samples whether an index is selected or not. Starting with the leftmost site, we take its probability distribution by tracing over the sites to the right by summing with the vector $\mathbf{1} = [1, 1]$, as shown in this diagram:
\begin{equation}
    \includegraphics[height=13.33ex]{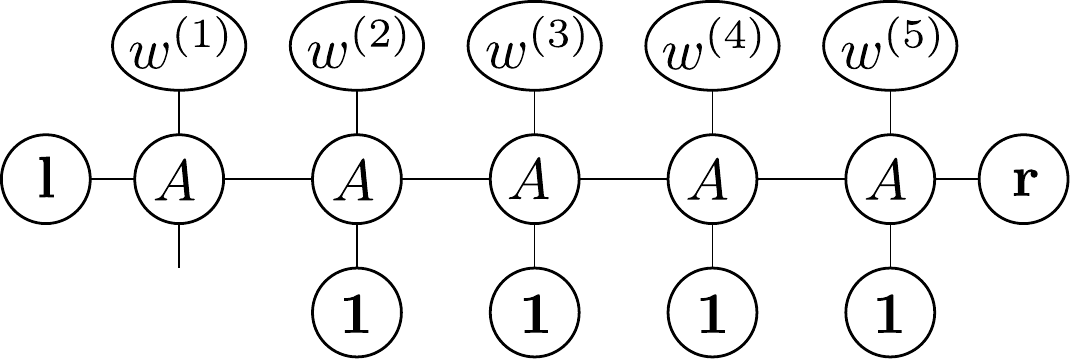}
\end{equation}
After generating a sample with probability $p(s_1)$ which is ratio of weights of selecting or not selecting the first state (leaving the site with fully determined probability distributions for `no' $\mathbf{n} = [1, 0]$ or for `yes' $\mathbf{y} = [0,1]$), we move on to the second site to determine it's conditional probability $p(s_2|s_1)$. The process is repeated. For example, after three steps, sampling `yes', `yes', `no' we end up with the following calculation of $p(s_4|s_1,s_2,s_3)$:
\begin{equation}
    \includegraphics[height=13.33ex]{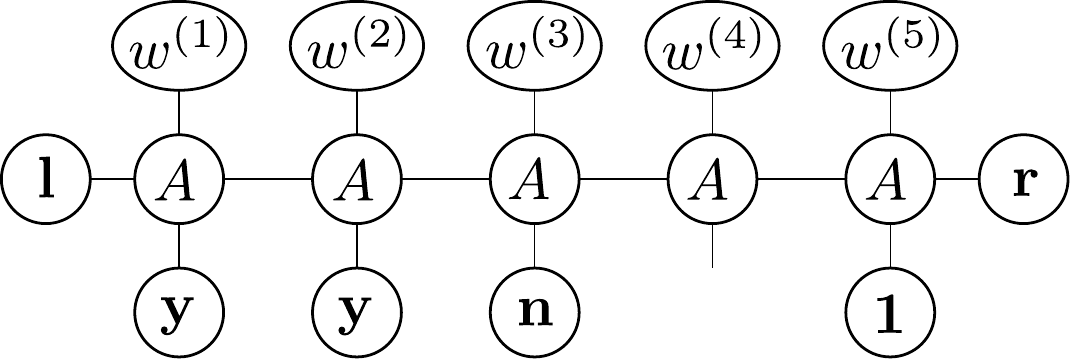}
\end{equation}
At the end we get the total probability distribution that we wanted,
\begin{multline}
   p(s_1) p(s_2|s_1) p(s_3|s_1,s_2) \dots p(s_D|s_1,\dots,s_{D-1}) \\
   = p(s_1,\dots,s_D).
\end{multline}

\begin{figure}[t]
\includegraphics[width=0.95\columnwidth]{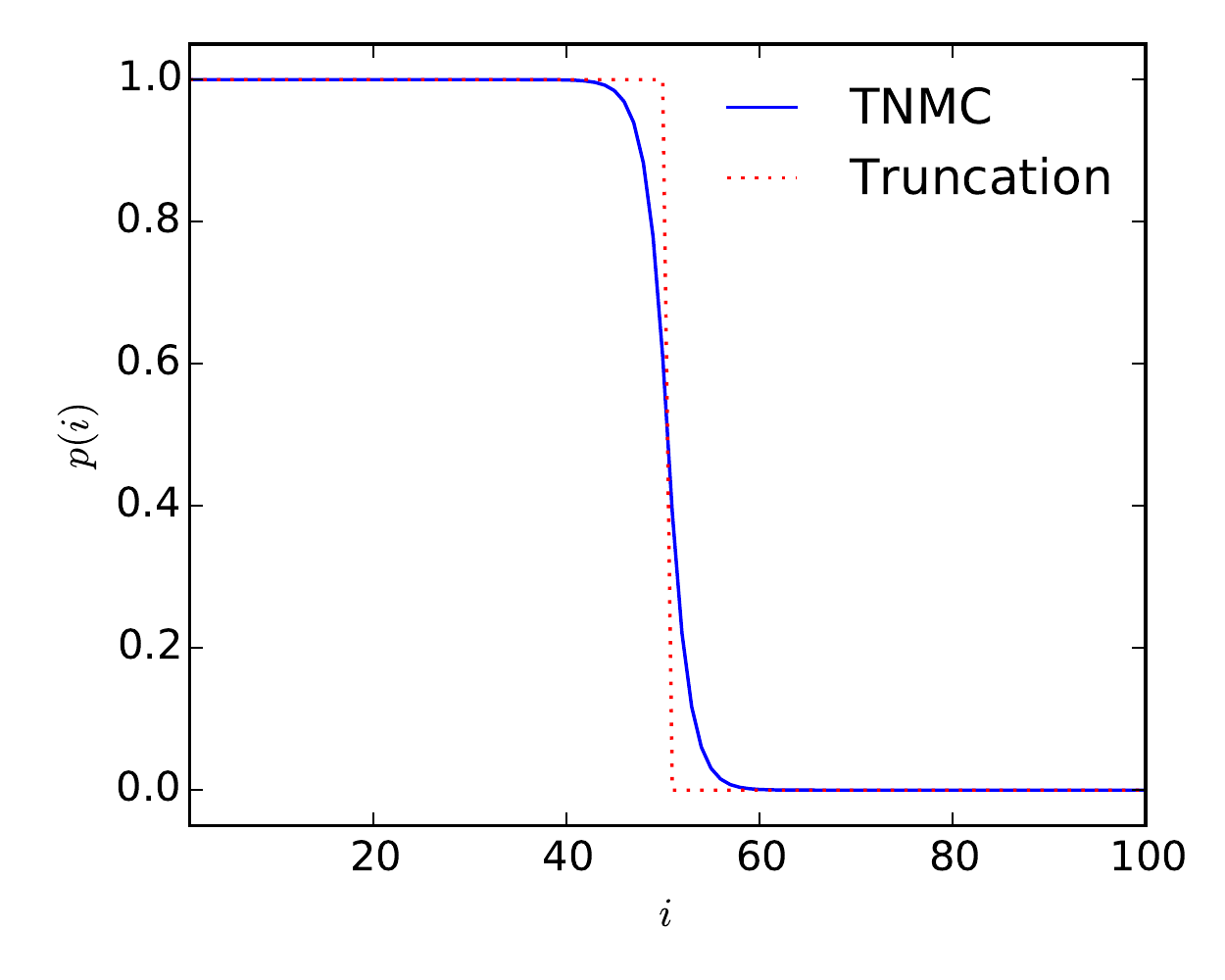}
\caption{The sampling distribution for singular values $S_{i}^2 = 2^{-i}$, when selecting 50 of 100 possible Schmidt vectors (i.e. $D=100$ and $D^{\prime}=50$). With high probability the largest are selected, while the tails of the distribution are sampled only occasionally. \label{fig:sampling}}
\end{figure}

The final thing that is required is the probability of selection of any individual Schmidt vector, $p(s_i)$. This is achieved simply by summing over all other configurations, like in $p(s_1)$ above. Below we calculate $p(s_4)$.
\begin{equation}
    \includegraphics[height=13.33ex]{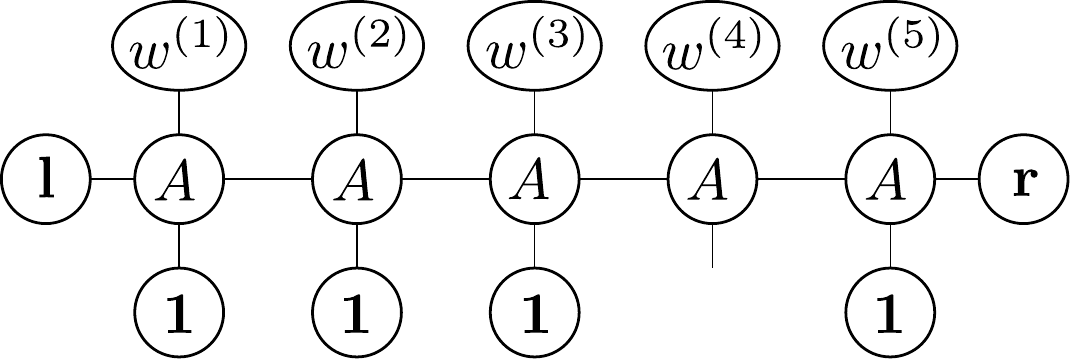}
\end{equation}
Each individual selected vector $i$ must be multiplied by the inverse of its appearance rate, $1/p(s_i)$, in order to create an unbiased Monte Carlo technique and to satisfy \Eqn{identity}. In \Fig{fig:sampling}, we show the sampling distribution $p(s_i)$ for a distribution of singular values $S_{i}^2 = 2^{-i}$. One notes that the sampling distribution looks very similar to the Fermi-Dirac distribution --- this is precisely because it \emph{is} the Fermi-Dirac distribution with constrained total particle number $D$ and appropriate energies $-k_B T \log(S_i^2)$.

The above steps can be performed in three simple `sweeps' of the matrix-product probability distribution. Care must be taken to avoid overflow and underflow, even of double-precision numbers, as the product in \Eqn{sampling_appendix} can grow \emph{very} large or small even with moderate $D$. Code for performing this algorithm in the \emph{Julia} programming language is provided freely and is available for download from github at \texttt{http://github.com/andyferris/sampling}.

\subsection{The Ising model as a tensor network state}

The thermal states of classical statistical models can generically be written in a tensor product form. We specialize here to the case of 2-body Hamiltonians acting on a square lattice, 
\begin{equation}
   H(\mathbf{s}) = \sum_{<i,j>} h(s_i,s_j),
\end{equation}
where $\mathbf{s} = [s_1,s_2,\dots]$ denotes a particular configuration. We wish to calculate the probability distribution
\begin{multline}
    p(\mathbf{s}) = \frac{\exp(-H(\mathbf{s})/k_B T)}{Z} \\
    = \frac{\exp\left(\sum_{<i,j>} h(s_i,s_j) / k_B T \right)}{Z} \\
    = \frac{\prod_{<i,j>} \exp\bigl(h(s_i,s_j)/k_B T\bigr)}{Z},
\end{multline}
and the related partition function $Z$
\begin{equation}
  Z = \sum_{\mathbf{s}} \prod_{<i,j>} \exp\bigl(h(s_i,s_j) / k_B T\bigr)
\end{equation}
To simplify the notation, we write the 2-body transfer matrix,
\begin{equation}
    T(s_1,s_2) = \exp\bigl(h(s_i,s_j) / k_B T\bigr),
\end{equation}
such that
\begin{equation}
  Z = \sum_{\mathbf{s}} \prod_{<i,j>} T(s_i,s_j) .
\end{equation}
For the Ising model in \Eqn{isinghamiltonian}, the transfer matrix is
\begin{equation}
    T(\cdot,\cdot) = \left[ \begin{array}{cc} \exp(J/k_B T) & \exp(-J/k_B T) \\ \exp(-J/k_B T) & \exp(J/k_B T) \end{array} \right].
\end{equation}

To express $Z$ as a tensor network, it is convenient to introduce a generalized Kronecker-$\delta$ function (also called a `copy tensor')
\begin{equation}
    \delta_{i,j,k,\dots} = \delta_{i,j} \delta_{i,k} \dots \;,
\end{equation}
which is zero unless $i=j=k=\dots$. It is drawn as a simple dot in tensor network diagrams, like
\begin{equation}
   \includegraphics[height=11ex]{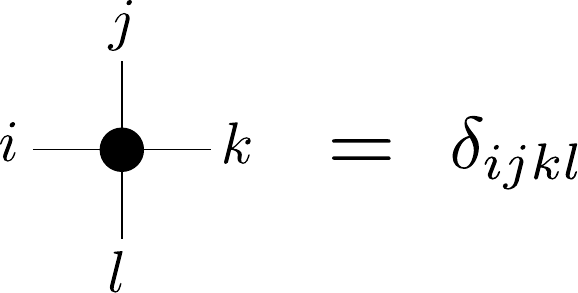}
\end{equation}

\begin{figure}[t]
  \includegraphics[width=0.6\columnwidth]{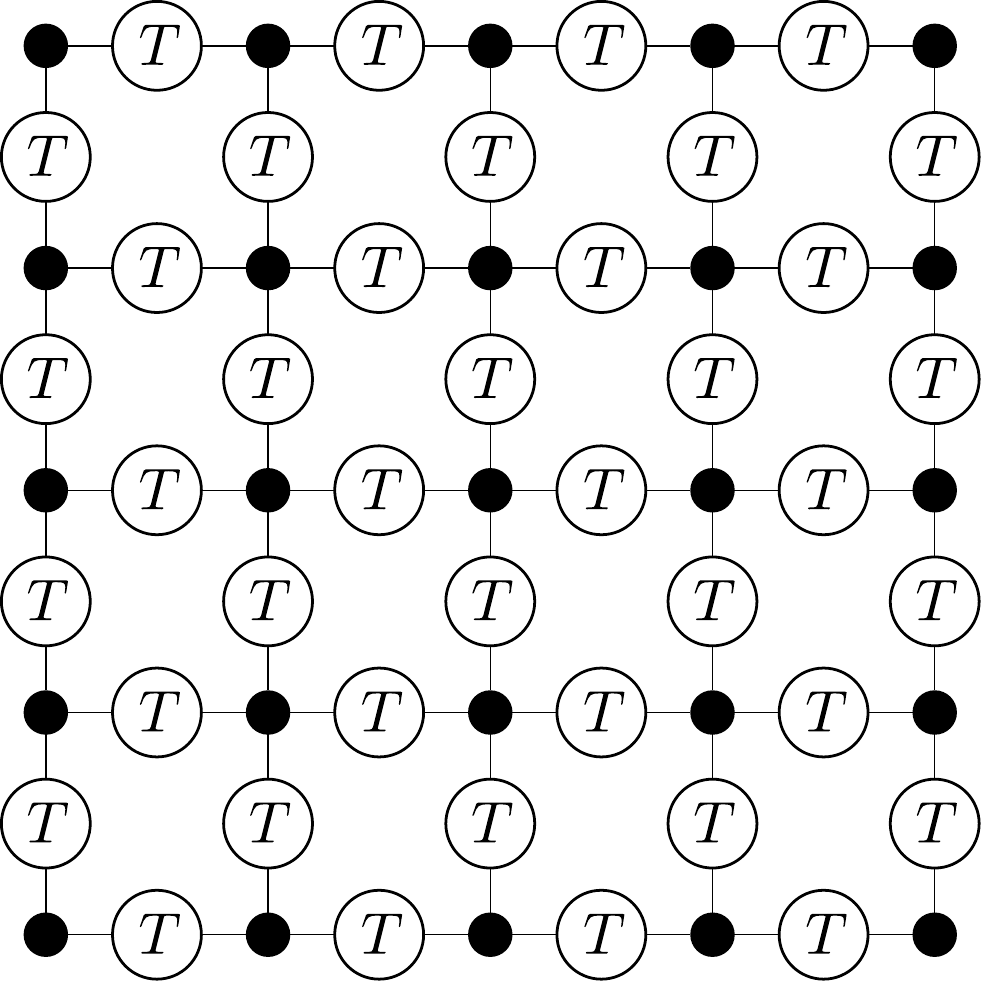}
  \caption{The partition function of a statistical system on an open square lattice with nearest-neighbour two-body interactions. \label{fig:partition}}
\end{figure}

This lets us easily `stitch' together the transfer matrices onto whatever lattice we like. The partition function on a square lattice is shown in \Fig{fig:partition}. To convert this to a standard `vertex' model, we make the following identification
\begin{equation}
    \includegraphics[height=15ex]{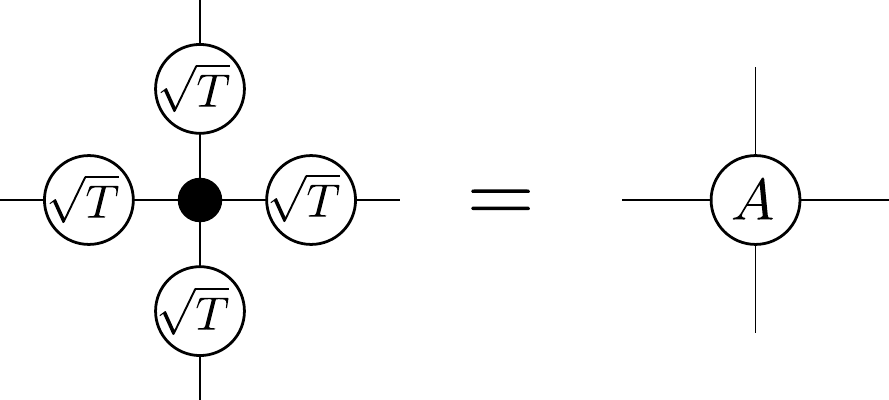}
\end{equation}
which leads to the structure in \Fig{fig:2dpartition}~(a), given the appropriate modifications to the tensors on the edges and corners. Extra bonds can be `opened up' at lattice sites for the purpose of measuring local observables, such as the magnetization in the centre of the lattice.
\begin{equation}
    \includegraphics[height=15ex]{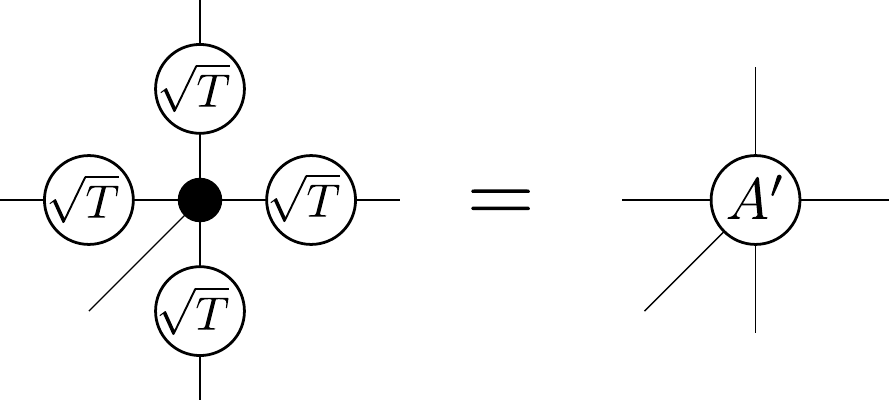}
\end{equation}

\end{document}